\newcommand{\di}{\partial}
\documentstyle[12pt]{article}
\begin{document}
\input{psfig}

\title{Worldsheet formulations of gauge theories and gravity}
\author{Michael P. Reisenberger\\
	Institute for Theoretical Physics\\
	University of Utrecht, P. O. Box 80 006\\
	3508 TA Utrecht, The Netherlands}
\maketitle

\begin{abstract}
The evolution operator for states of gauge theories in the graph representation
(closely related to the loop representation) is formulated as a weighted sum
over
worldsheets interpolating between initial and final graphs. As examples,
lattice $SU(2)$ BF
and Yang-Mills theories are expressed as worldsheet theories, and
(formal) worldsheet forms of several continuum $U(1)$ theories are given.

It is argued that the world sheet framework should be ideal for representing
GR, at least euclidean GR, in 4 dimensions, because it is adapted to both the
4-diffeomorphism invariance of GR, and the discreteness of 3-geometry
found in the loop representation quantization of the theory.
However, the weighting of worldsheets in GR has not yet been
found.
\end{abstract}

\section{Introduction}

This is a talk presented at the 7th Marcel Grossmann Meeting, held
at Stanford in July '94. It is a status report on a spacetime worldsheet
formulation of quantum gauge
theories (including GR) which the author has been developing\footnote{
Results of this program restricted to $U(1)$ theories were reported at
the conference {\em New Directions in General Relativity}, Maryland May 1993,
and at the {\em Midwest Relativity Meeting}, Detroit, November 1993.}
Most proofs are omitted.\footnote{A manuscript, including proofs, is in
preparation.}

\begin{figure}
\centerline{\psfig{figure=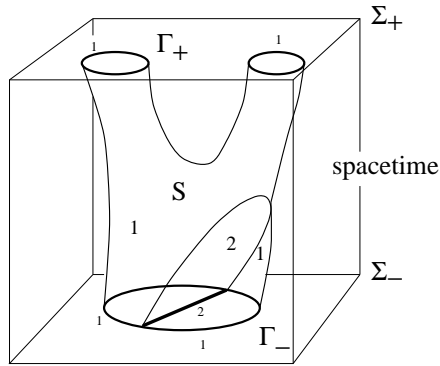,height=5cm}}
\caption{In the worldsheet formulation the amplitude,
$U^{\Gamma_+}{}_{\Gamma_-}$, to evolve from graph state $|\Gamma_-\rangle$
on the initial hypersurface $\Sigma_-$, to graph state $|\Gamma_+\rangle$
on the final hypersurface $\Sigma_+$ is given by a sum over interpolating,
possibly branched, 2-surfaces $S$. Both the edges, and vertices
of valency $r>3$ of the graph and the simple
components, and lines of intersection with $r>3$ of $S$ carry irreducible
representations of the gauge group.}

\label{worldsheet}
\end{figure}

The idea of this reformulation of gauge theories is to express the evolution
operator for states in the loop representation
\cite{Gambini84} \cite{Gambini86} \cite{RSloops} \cite{RSloops90} as a sum over
2-surfaces interpolating between initial and final loops (see Fig.
\ref{worldsheet})
The overcompleteness of the loop basis complicates the situation somewhat.
It is easier to work with the closely related graph basis (defined in
Section \ref{derivation}), associated with graphs whose edges, and vertices,
carry gauge
group representations, instead of loops. In the worldsheet formulation
described in the present paper the evolution operator in the graph
representation is expressed as a sum over 2-surfaces interpolating between
initial and final graphs. See Fig. \ref{worldsheet}.

The chief motivation for developing this version of loop quantization is that
it
makes the 4-diffeo invariance of generally covariant theories manifest: the
weight
of each interpolating worldsheet $S$ depends only on the 4-diffeo equivalence
class of $S$. Thus the worldsheet integral becomes a sum over 2-knots (or, more
precisely, 2-tangles).

\begin{figure}
\centerline{\psfig{figure=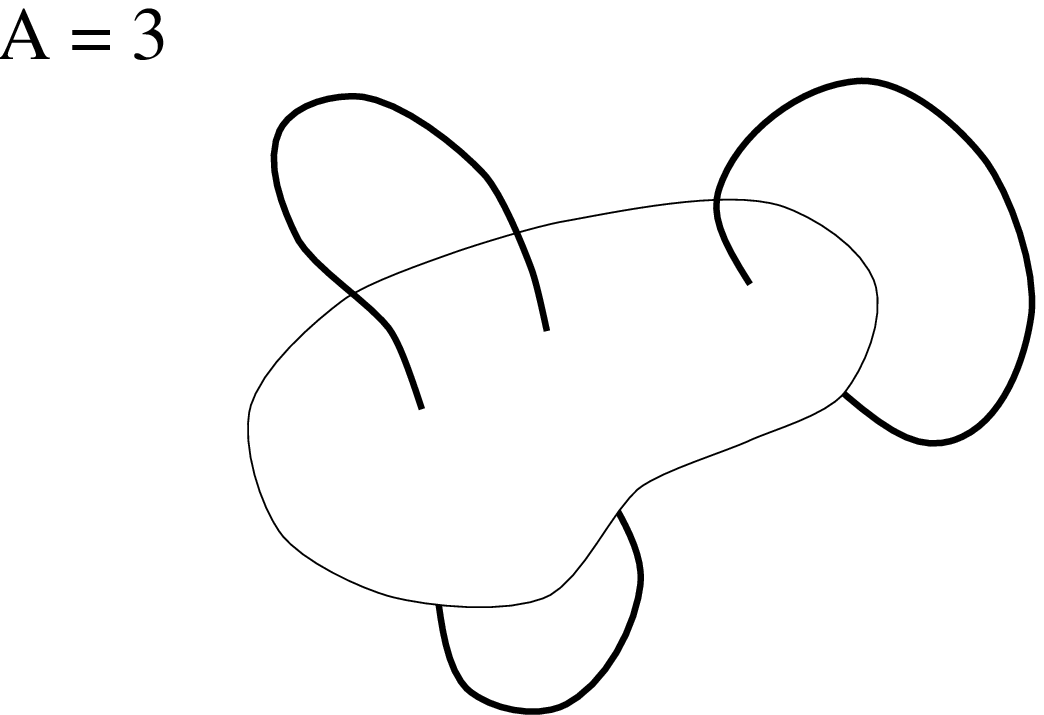,height=4cm}}
\caption[]{ According to \protect\cite{discarea1} both the area operator of a
surface
$\sigma$ and the volume operator of a region of space, $\mu$, are block
diagonalized
in the loop basis (and, by a simple extension, in the graph basis)
in the sense that they act only within the subspaces of loops, or graphs,
having the same shadow, i.e., those having the same edges, differing only in
the
structure of their intersections. On loop states labeled by collections,
$\gamma$, of
smooth closed loops the area of $\sigma$ is \newline
$A = \frac{\sqrt{3}}{2}\ (\mbox{Plank length})^2\times
(\#\ \mbox{intersections of $\sigma$ with $\gamma$})$.
\newline
On the full state space the spectra are more complicated, but still discrete.}
\label{area_volume}
\end{figure}

		A second reason for pursuing a worldsheet formulation is that it
may reveal
a microstructure of spacetime mandated by GR. In canonical loop quantization it
has been found \cite{discarea1} that the quantization of the classical
area observable of a 2-surface essentially counts the (unsigned) number of
crossings
of loops through the 2-surface.
of loops (see Fig. \ref{area_volume}).\footnote{In a paper which appeared after
MG7,
\cite{discarea2}, Rovelli
and Smolin re-examine the quantization of area and volume using, instead of
the loop representation, the slightly different graph representation described
in the present paper. The discreteness found
in \cite{discarea1} is confirmed. Furthermore, it is argued
in \cite{discarea2} that if matter fields are used to define the spacetime
2-surfaces and
3-surfaces which the area and volume refer to, the operators
become observables in the physical state space with unchanged spectrum,
so that these spectra are, in fact, physical predictions of the quantization of
GR used.}

Thus the loop basis states can be thought of as eigenstates of the 3-geometry
of
space,
and the 3-geometries in the spectrum are ``discrete" and determined by the
loops.
One expects that the sum over worldsheets in the evolution operator represents
a
sum
over discrete {\em spacetime} geometries specified by the worldsheets (in
which,
for instance, the area of a spacetime 2-surface is determined by the number of
its intersection points with the loop worldsheet).

\begin{figure}
\centerline{\psfig{figure=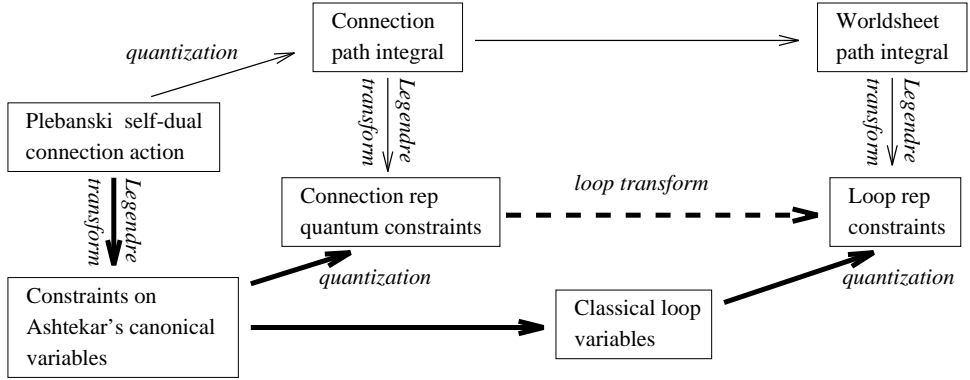,height=5cm}}
\caption{Maps connecting various forms of classical and quantum GR. The bold
lines mark the ``beaten paths" thought to be well understood. The loop
transform, represented by a bold, dashed arrow, is becoming better understood
\protect{\cite{Ashtekar93}}. The author proposes to establish the top chain
of links, from the classical Plebanski action to a worldsheet path integral,
and then down to the canonical loop representation formulation of GR.}
\label{maps}
\end{figure}

A final, concrete, motivation is to check the loop Hamiltonian constraint. By
transforming
GR from a spacetime connection form directly to a worldsheet form and then
going,
by a Legendre transformation, to the canonical loop formulation the Hamiltonian
constraint
can be derived in a new way (see Fig. \ref{maps}). The derivation of the
Hamiltonian
constraint
has, so far, been the least ``natural" step in the loop quantization program.

		The features of a worldsheet formulation of GR have been
described tentatively because
no such formulation is known as yet. However, the author has developed, 1), a
worldsheet form
of $U(1)$ gauge theories in the continuum and, 2), a world sheet form of
$SU(2)$ gauge theories formulated on an arbitrary, not necessarily
hypercubic, lattice. Using these results, explicit worldsheet actions have been
found for regulated,
euclidean, compact EM, $U(1)$ F-F dual theory, a $U(1)$ analogue
of GR,
$SU(2)$ BF theory and $SU(2)$ lattice YM theory. Euclidean GR can be thought of
as an $SU(2)$
gauge theory of the left-handed spin connection \cite{Plebanski77}
\cite{CDJ}. The corresponding worldsheet formulation is being
developed by the author.

		Before describing the above results in a little more detail,
let's outline how the
$SU(2)$ lattice worldsheet formulation is derived.

\section{Derivation of the worldsheet formulation} \label{derivation}

\begin{figure}
\centerline{\psfig{figure=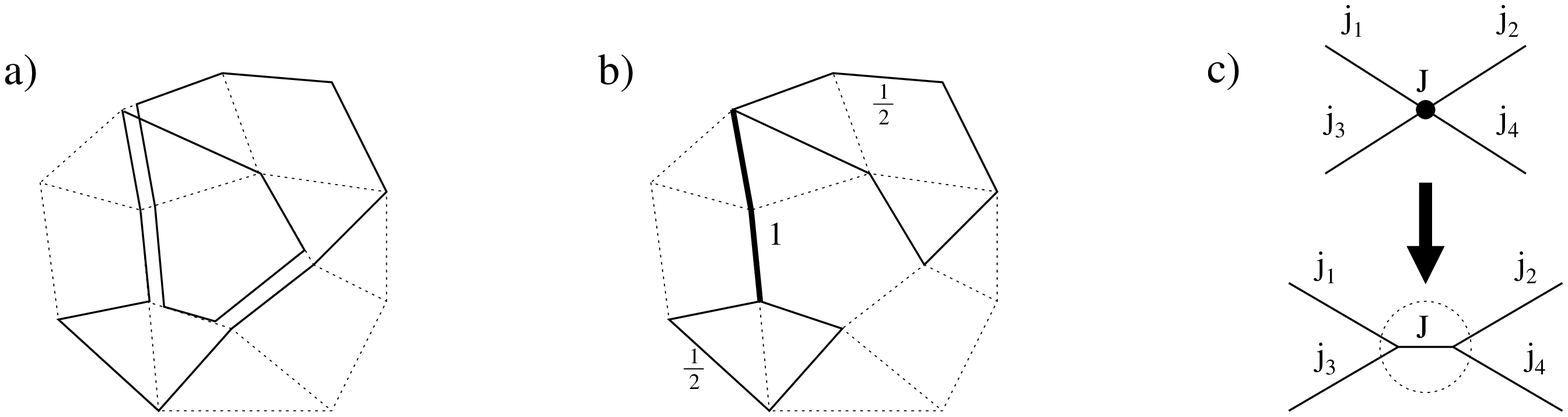,height=3.6cm}}
\caption{Panel a) illustrates a loop state. A closed line travels along
the links of the lattice, possibly returning to the same link. Each line
segment along a link represents a factor $U^{(\frac{1}{2})}(g)$, the
fundamental representation matrix of the link variable $g$, with
one index associated with each end of the segment. The connections
of the segments represent the contractions of the indices of the
$U^{(\frac{1}{2})}$s.
\newline
Panel b) illustrates a graph state. A closed graph lies on the links of the
lattice, with at most one line segment on each link. The line segments each
represent a factor $U^{(j)}(g)$, where the (non-zero) spin $j$ is given
by the color of the line. Note that the states shown in a) and b)
are not equivalent, although a) is present in an expansion of b) into
loop states.
\newline
Panel c) illustrates the expansion of a colored 4-valent vertex into a
colored trivalent graph,
which is used in the definition of graph states. }
\label{bases}
\end{figure}

The most obvious basis of
$SU(2)$ gauge
invariant states on a lattice is the loop basis, given in the link
representation
by the Wilson loops $W_\gamma$ (see Fig.~\ref{bases}a).
\begin{equation} \label{Wilson_loop}
\langle\{g_l\}_{l\in \mbox{\scriptsize links}}|\gamma\rangle =
\prod_{l\in\gamma} g_l = W_\gamma
\end{equation}
(A given link, $l$, may appear in the loop, $\gamma$, several times). This
basis is
overcomplete. There is a convenient, complete, and linearly independent
basis closely related
to the loop basis. It will be called the ``graph" basis here. Its elements are
labeled by graphs,
$\Gamma$, consisting of edges labeled or ``colored" by the spins
$j\in\{\frac{1}{2},1,\frac{3}{2},2,...\}$ of non-unit irreducible
representations of SU(2), and vertices, of arbitrary valency $r$,
colored by $r-3$ $j$'s.(see Fig \ref{bases}b).
The function, $\Gamma(\{g_l\}) \equiv \langle\{g_l\}_{l\in
\mbox{links}}|\Gamma\rangle$,
representing the graph basis state $|\Gamma\rangle$ in the link representation
is given by a
factor $U^{(j_e)m_+ m_-}(g_e)$ for every edge $e\in\Gamma$ (with $g_e =
\prod_{l\in e} g_l$ ordered along $e$)
contracted on $m$'s with a factor $({}^{j_1}_{m_1} {}^{...}_{...}
{}^{j_r}_{m_r})_{J_1...J_{r-3}}$ for
each ($r$-edge) vertex. $U^{(j)}(g)$ is the spin $j$ representation of
$g \in SU(2)$, and
$({}^{j_1}_{m_1} {}^{...}_{...} {}^{j_r}_{m_r})_{J_1...J_{r-3}}$ is
the generalized Wigner coefficient defined in \cite{Yutsis62}.

$({}^{j_1}_{m_1} {}^{...}_{...} {}^{j_r}_{m_r})_{J_1...J_{r-3}}$ is
obtained by expanding the vertex into a tree graph with $r$ external legs
and $r-2$ oriented, trivalent vertices connected by
$r-3$ internal edges carrying spins $\{J_i\}$ and link variables
$\{g_i = {\bf 1} \}$, and contracting with $3-j$ symbols at each 3-vertex.
At each vertex a conventional choice must be made of which
tree graph to use in the expansion. The internal spins $\{J_i\}$
label distinct states within the same graph basis. Different choices
of trees, on the other hand, lead to different graph bases.\footnote{The
colored graph $\Gamma$ with its vertices expanded into colored trivalent
trees is a ``spin network". See \cite{Kauffman91} \cite{Penrose69}.} The
expansion
of vertices is illustrated in Fig. \ref{bases}c.

The graph basis, as well as several ideas used here, are found, at least
implicitly, in Ooguri's paper \cite{Ooguri92}. Since MG7 two papers have
appeared discussing the graph basis in detail: \cite{discarea2} and
\cite{baez94}.

$SU(2)$ graph states can be
expanded in loop basis states by writing the spin $j$ representation matrices
$U^{(j)}$ as symmetrized outer products of $2j$ factors of the corresponding
fundamental representation matrix $U^{(\frac{1}{2})}$ \newline
($U^{(j)}_{m_+}{}^{m_-} = U^{(\frac{1}{2})}_{(M_{+,1}}{}^{(M_{-,1}}
U^{(\frac{1}{2})}_{M_{+,2}}{}^{M_{-,2}}\ ...\
U^{(\frac{1}{2})}_{M_{+,2j})}{}^{M_{-,2j})}$,
$m_\pm = \sum_{n=1}^{2j} M_{\pm,n}$).
For example,

\centerline{\psfig{figure=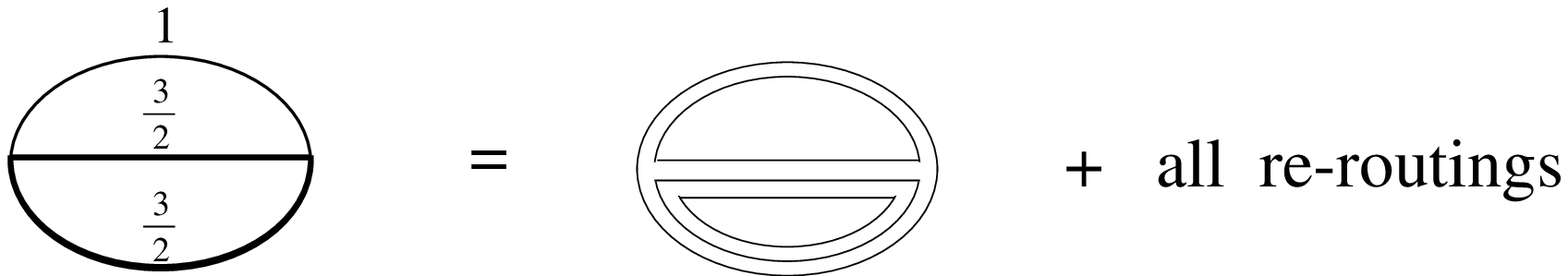,height=2cm}}

The graph states are orthogonal with respect to the simple, positive definite
inner product\footnote{
It is {\em not} claimed that $\langle\ \ |\ \ \rangle$ is the physical inner
product, according to
which the Hamiltonian and observables are hermitian. It is simply a useful
inner product.
Unfortunately the bra-ket notation does not accomodate a multiplicity of inner
products well.}
\begin{equation}\label{inner_product}
\langle\theta|\psi\rangle = \prod_{l\in\mbox{\scriptsize links}}
\int_{SU(2)} \theta^*(\{g_l\})\ \psi(\{g_l\})\ dg_l,
\end{equation}
where $dg$ is the invariant, or Haar, measure on the gauge group $SU(2)$, with
$\int_{SU(2)}dg = 1$.
The graph states are normalized by multiplying by a factor $\sqrt{2j + 1}$
for each edge, including the fictitious ones in the expansion of vertices
into trivalent trees. From now on, the graph states will be assumed to be
normalized.

The coefficient of each graph state in the graph expansion of an
arbitrary state is easily found:
\begin{equation} \label{graph_expansion}
|\psi\rangle = \sum_\Gamma |\Gamma\rangle\langle\Gamma|\psi\rangle.
\end{equation}
{}\footnote{Graph bases can be defined for the state space of any
gauge theory in which the irreducible representations of the
gauge group provide a complete basis of functions on the group manifold.
By the Peter-Weyl theorem this is true for all compact Lie groups.

In a $U(1)$ gauge theory the edges of the graphs defining the graph basis
are labeled by $n \in {\bf Z},\ n\neq 0$ (because the irreps of $z =
e^{i\,\theta} \in U(1)$ are $\{ z^n | n\in {\bf Z}\}$), and the vertex factors
are $1$. The decomposition of graph states into loop states is possible
as long as all irreps are contained in outer products of the fundamental
rep.}

Note that states in the loop representation of Gambini and Trias, and
Rovelli and Smolin \cite{Gambini84} \cite{Gambini86} \cite{RSloops}
\cite{RSloops90}
are functions of loops defined by $\psi[\gamma] = \langle
W_\gamma^*|\psi\rangle$. $\psi[\gamma]$ is {\em not}, in general, the
coefficient of $|\gamma\rangle$ in an expansion of $|\psi\rangle$
on the loop basis. In other words, $|\psi\rangle \neq \sum_\gamma|\gamma\rangle
\langle W_\gamma^*|\psi\rangle$. The great advantage of the graph
representation
is that the analogous equation, (\ref{graph_expansion}), does hold.
The quantity we want to represent as a sum over worldsheets is the evolution
operator $\hat{U}$.
(In diffeo invariant theories this is just the projector to the physical state
space). It maps the state, $|\psi_-\rangle$, of the field on the initial
time hypersurface, $\Sigma_-$, to its evolved image, $|\psi_+\rangle$,
on the final time hypersurface, $\Sigma_+$.
\begin{equation}
|\psi_-\rangle = \hat{U}|\psi_+\rangle.
\end{equation}

Expanding both $|\psi_+\rangle$ and $|\psi_-\rangle$ in graph basis states
leads
to
\begin{eqnarray}
|\psi_+\rangle & = & \sum_{\Gamma_+}\sum_{\Gamma_-}
|\Gamma_+\rangle\langle\Gamma_+|\hat{U}|\Gamma_-\rangle
\langle\Gamma_-|\psi_-\rangle \\
& \equiv &  \sum_{\Gamma_+}\sum_{\Gamma_-}
|\Gamma_+\rangle\ U^{\Gamma_+}{}_{\Gamma_-}\ \langle\Gamma_-|\psi_-\rangle.
\end{eqnarray}
$U^{\Gamma_+}{}_{\Gamma_-}$ is the coefficient of $|\Gamma_+\rangle$ in the
evolved image of
$|\Gamma_-\rangle$.

In the link representation (in which states are written as functions of the
set of link variables $\{g_l\}$) the evolution operator is given by a
Feynman ``path integral" over link variables belonging to links in the
spacetime
region ${\cal M}$, bounded by $\Sigma_+$ and $\Sigma_-$:
\begin{equation}
U^{\{g_{l_+}\}}{}_{\{g_{l_-}\}} = \prod_{l \in {\cal M} - \Sigma_+\cup\Sigma_-}
\int_G dg_l\ e^{i\,I}
\end{equation}
where the action $I(\{g_l\})$ depends on the $g_l$ of all $l\in \bar{\cal M}
= {\cal M} \cup \Sigma_+ \cup \Sigma_-$.\footnote{
There is a subtlety here. In general there will be terms in $I$ which
depend only on links in $\di {\cal M} = \Sigma_+\cup \Sigma_-$. For instance,
in lattice YM theory there will be contributions from plaquettes in
$\di {\cal M}$. The operator $\hat{U}(t_+, t_-)$ for evolution from time $t_-$
to $t_+$ defined above does not satisfy $\hat{U}(t_+, t_-) =\hat{U}(t_+, t)
\hat{U}(t, t_-)$ as it should, because the boundary contributions to $I$ on
$\Sigma_t$ are counted twice. The nicest way to avoid this problem is to define
states on ``interstitial 3-surfaces" formed by the 3-faces of the dual lattice.
This requires a considerable change in the formalism, without substantially
changing the results. Thus the whole issue is ignored in this sketch.}

Translating this into the graph representation yields
\begin{eqnarray}
U^{\Gamma_+}{}_{\Gamma_-} & = & \langle\Gamma_+|\hat{U}|\Gamma_-\rangle \\
	& = & \prod_{l\in \bar{\cal M}} \int_G dg_l\ \Gamma_+^*\ e^{i\,I}
\ \Gamma_-.          \label{evolution}
\end{eqnarray}

Gauge invariance implies that the Feynman weight, $e^{i\,I}$, can always be
written as a sum of
products of factors
associated with plaquettes.\footnote{To prove this begin by noting that gauge
invariance
forces $e^{iI}$ to be a sum of ``graph states" $\Gamma(\{g_l\})$ defined on
spacetime instead
of 3-space.}
\begin{eqnarray}
e^{i\,I} & = & \sum_{\{j_p\}}\ \ [\ \sum_{\{m\}} c_{\{j_p\},\{m\}}
\prod_{p\in\mbox{\scriptsize plaquettes}} f_p\ ]
\label{jp_decomposition} \\
f_p & = & U^{(j_p)}(g_{l_1})\otimes U^{(j_p)}(g_{l_2})\otimes ...\otimes
U^{(j_p)}(g_{l_n}).
\end{eqnarray}
The links $l_1, ..., l_n$ make up the boundary of $p$.
$\{m\}$ in the sum (\ref{jp_decomposition}) refers to the indices of the
matrices $U^{j_p}_{m_{l+}}{}^{m_{l-}}(g_l)$ in the outer
product $f_p$. ($m_{l-}$ is associated with the beginning of the link $l$,
and $m_{l+}$ with the end).

In a given gauge theory there is a great deal of lattitude in the choice of
coefficients $c$ in (\ref{jp_decomposition}). Theories are ``local" if
$c$ can be chosen to be a product of factors $c_s$, each correlating only
the link variables $g_l$ of links touching that site. The $c_s$ are to depend
only on the $j$s of plaquettes touching $s$, and on indices $m$ associated with
$s$.

This notion of locality seems fairly reasonable in the link formulation.
It holds, in particular, for $SU(2)$ BF and YM theory.\footnote{
One can widen the notion of local theories somewhat by requiering only that
$c$ be a product of factors associated with ``local blocks": disjoint,
finite collections of adjacent sites. This may be necessary to accomodate
GR.}
Moreover, if a gauge theory is local in this sense, in its worldsheet form
the worldsheet will only interact locally.

\begin{figure}
\centerline{\psfig{figure=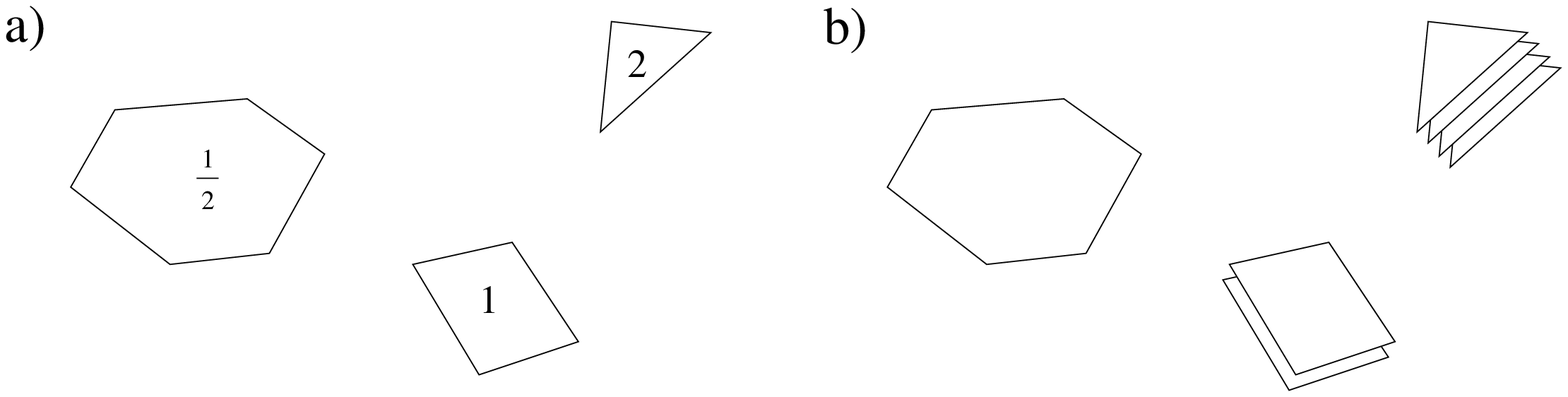,height=3cm}}
\caption[]{Panel a) shows a term in the sum (\protect{\ref{jp_decomposition}})
in the colored
surface formulation. The numbers on the polygons, which represent occupied
plaquettes, are the
spins, $j_p$, they carry.

Panel b) shows the elementary surface picture of the same term. Each plaquette
in a) is replaced
by $2j_p$ copies of itself.}
\label{patches}
\end{figure}

In the terms of (\ref{jp_decomposition}) each plaquette carries a particular
irrep $j_p$.
In fact, each term in the sum over $\{j_p\}$
can be visualized as a collection of patches of surface, namely the occupied
plaquettes $\{ p | j_p \neq 0 \}$, labeled or ``colored" by their
$j_p \in \{\frac{1}{2}, 1, \frac{3}{2}, 2, ...\}$. (See Fig. \ref{patches}a.)
The sum over $\{j_p\}$ in (\ref{jp_decomposition}) can thus be seen as a sum
over surfaces (with boundaries inside $\cal M$, in general) and the coloring
of these surfaces.

Expanding the integrand in (\ref{evolution}) according to
(\ref{jp_decomposition})
and carrying out the integral over the link variables leads to an expression
for the evolution operator a sum over colored surfaces.

Let me stop to mention here a slightly different form of the expansion
(\ref{jp_decomposition}),
which leads to a slightly different expression for the evolution operator.
Each of the factors $U^{(j_p)}$ in (\ref{jp_decomposition}) can be expanded
as an outer product of $2j_p$ factors of
$U^{(\frac{1}{2})}$, symmetrized on first and second indices separately. This
expanded form of the terms in (\ref{jp_decomposition})
can be visualized by replacing each occupied plaquette in the colured surface
picture by a stack of $2j_p$ copies of itself. (See Fig.~\ref{patches}b). Each
edge $l$ of the
colored plaquette, which represents $U^{(j_p)}(g_l)$, is thus replaced by
$2j_p$
edges, each representing a factor of $U^{(\frac{1}{2})}(g_l)$. This is the
``elementary surface" picture of the terms in (\ref{jp_decomposition}).
Note that the expansion of spin $j_p$ edges into $2j_p$ spin $\frac{1}{2}$
edges, used to go from the colored surface picture to the elementary surface
picture, is exactly the same as that used to expand colored graph basis
states in loop basis states.

Back to the colored surface formulation.
Inserting (\ref{jp_decomposition}) in (\ref{evolution}) yields an
expression for the evolution amplitude which is both a sum over
$\{j_p\}$ and an integral over the link variables $\{g_l\}$.
The integration over the $\{g_l\}$ of each term in the sum over $\{j_p\}$ is
easily carried out.
The integral is
zero unless the surface, $\{ p |j_p \neq 0\}$, is continous and its edges on
$\di {\cal M}$ match up with the final and initial graphs, $\Gamma_+$ and
$\Gamma_-$, both in
space and in coloration.\footnote{
On a link, $l$, where $r>3$ surfaces meet the integration over $g_l$ leaves
$r-3$ discrete degrees
of freedom associated with the link. These are treated in complete analogy with
the vertex
degrees of freedom of graph basis states. The link, $l$, is expanded to a
trivalent ``tree
surface", i.e., $(\mbox{a trivalent tree graph})\times l$, consisting of $r-3$
2-cells carrying
spins $J_1,...,J_{r-3}$, and the factor \newline
$1 = (2J_1 + 1)...(2J_{r-3} +1) \sum_{\{m\},\{n\}}
({}^{j_1}_{m_1} {}^{...}_{...} {}^{j_r}_{m_r})_{J_1...J_{r-3}}
U^{(j_1)m_1 n_1}...U^{(j_r)m_r n_r}
({}^{j_1}_{n_1} {}^{...}_{...} {}^{j_r}_{n_r})_{J_1...J_{r-3}}$\newline
is inserted in the Feynman weight.
The expansions of the links is taken to be compatible with those of the
vertices of the graphs on $\di {\cal M}$. Then, in order that the integral over
$\{g_l\}$ not vanish
the coloration of the tree surfaces must match those of the tree expansions of
the vertices
of the graphs on the boundary.}
The evolution
amplitude can, therefore, be written as a sum over continuous (but possibly
branched)
surfaces interpolating between $\Gamma_-$ and $\Gamma_+$. For such continuous
interpolating surfaces the integral over $\{g_l\}$ yields an outer
product of SU(2) invariant tensors (constructed from Wigner 3-j symbols)
which contract the indices of the coefficient $c$ in (\ref{jp_decomposition})
to give the weight $e^{i\,I(S)}$ of $S$ in the sum. Thus
\begin{equation} \label{surf_sum}
U^{\Gamma_+}{}_{\Gamma_-} = \sum_{\{S|\di S = \Gamma_+\cup\Gamma_-\}}
e^{i\,I(S)}.
\end{equation}
The symbol $S$ stands for both the surface and it's coloration.

\begin{figure}
\centerline{\psfig{figure=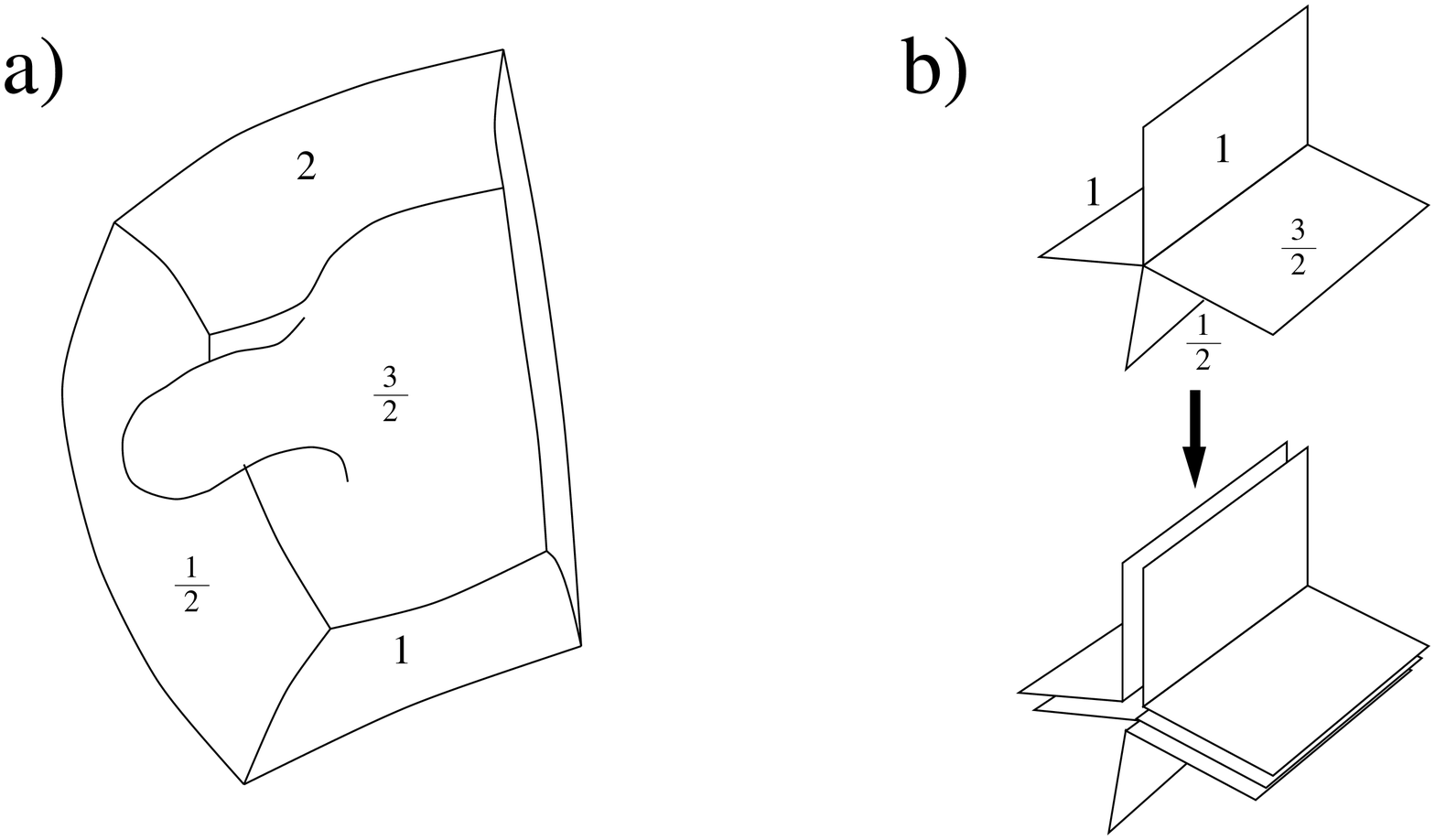,height=5cm}}
\caption[]{ Panel a) illustrates that $S$ in
 (\protect\ref{surf_sum}) is the union
of monochrome simple surfaces bounded by lines on which $3$ or more surfaces
meet. The simple components can have nontrivial topology and need not be
orientable in SU(2) theories. There can also be transverse intersections
of simple components at points. Note that at most one simple component
traverses a given plaquette.

Panel b) illustrates the polygon inequalities at lines of intersection. One way
of
stating these is that there must exist a routing of simple surfaces through the
line of
intersection so that $2j_\sigma$ such surfaces enter (or exit) the line of
intersection
along each simple component $\sigma$ of $S$ meeting at the line, with none of
the surfaces
returning along the same component. }
\label{Surface}
\end{figure}

$S$ is the union of a collection of simple (unbranched) surfaces, $\sigma$,
bounded by
lines on which $3$ or more simple surfaces join.
Only those $S$ on which each simple component carries a single uniform color
$j_\sigma$ contribute to (\ref{surf_sum}).
(see Fig. \ref{Surface}a).
Furthermore the colors of simple components joining at a line must
satisfy polygon inequalities, i.e., the $j$'s must be the edge lengths of a
closed
polygon of integral perimeter.
See Fig. \ref{Surface}b for an illustration of an illuminating alternative
formulation of this
requirement.

Instead of colored surfaces one may work with elementary surfaces.
Proceeding in exact analogy to the derivation of the colored surface
formulation one finds an expression for the evolution amplitude from
the initial to the final graph basis state as a sum over continous,
elementary surfaces which interpolate between loops present in the
decomposition of $| \Gamma_+\rangle$ and $|\Gamma_-\rangle$ into loop
basis states. Unlike the colored surfaces the elemetary surfaces
can traverse a given plaquette several times.
The elementary surfaces are not branched, in the sense that there
are no lines on which $3$ or more non-coplanar surfaces join. However,
there can be branch cuts ending at branch
points, as occur on Riemann surfaces.

This completes the outline of the derivation of the world sheet formulation
of general $SU(2)$ gauge theories. For $U(1)$ theories the world sheet
formulation
can be derived analogously, on a lattice, or directly in the continuum
by a method which will not be discussed here.

\section{Worldsheet forms of some specific theories}

Let's first summarize the results for $U(1)$ theories.
In the conection formulation, which we take as our starting point,
these are theories of $U(1)$ fiber bundle geometries, described by a curvature,
or field
strength, 2-form $F$. In a neighborhood about any point not on a magnetic
monopole worldline one can find a 1-form potential $A$ such that
$F = d\wedge A$. The $U(1)$ bundle is not assumed trivial, so Dirac monopoles
are allowed, at least kinematically.

In the worldsheet formulation of $U(1)$ theories the evolution operator is a
sum, weighted
by the world sheet Feynman weight $e^{i\,I_s}$, over {\em oriented} 2-surfaces
interpolating
between initial and final graphs.

Oriented 2-surfaces can be added and multiplied by integers in the sense that
$\int_{m S_1 + n S_2}
= m\int_{S_1} + n\int_{S_2}$, so they form an infinite dimensional lattice,
even in the continuum.
The sum over surfaces in the elementary surface formula for the evolution
operators of
$U(1)$ theories are sums over this
lattice, with equal {\em a priori} weights for each lattice point. Whether the
sums are well
defined depends on the Feynman weight function of the particular theory being
considered.

For regulated Euclidean compact electrodynamics with connection action
$I_c [A] = \frac{1}{4}\int F_{\mu\nu} F_{\lambda\rho} \eta^{\mu\lambda}
\eta^{\nu\rho} d^4x$ the surface action is
\begin{equation}
I_s(S) = \frac{e^2}{\Lambda^2} \int_S n^2 d(\,area\,)  \label{EM_surf}
\end{equation}
where $\Lambda$ is the regulator length, $e$ is the fundamental electric
charge, and
$n(x)$ is the representation of $U(1)$ at $x$ on the surface $S$. In the
elementary
surface formulation the action is still (\ref{EM_surf}), with $n(x)$ the is the
number
of layers of elementary surface at $x$. It
is interesting that, besides the $n^2$ dependence, this is just the Nambu-Goto
action
of string theory. The summation over worldsheets appears to be well defined in
this theory.

Compact EM has a well known \cite{Banks77} phase transition from a strong
coupling phase,
in which electric charges are confined, to a weak coupling phase, in which
charges are
unconfined and the EM field consists of a collection of non-interacting
photons.
This transition is currently being studied in the worldsheet formalism.
Preliminary
indications are that the worldsheet action (\ref{EM_surf}) describes this
transition.
At the transition the elementary surface form of the worldsheet sum is
dominated by
spacetime filling worldsheets. Thus the $n^2$ factor in the the lagrangian
of (\ref{EM_surf}) becomes important at the transition.

	F~-~F dual theory, with $I_c[A] = k \int \epsilon^{\mu\nu\lambda\rho}
F_{\mu\nu} F_{\lambda\rho}\ d^4x
= k \int F\wedge F$ has the surface action
\begin{equation}
I_s(S) = -\frac{e^2}{4k}\ \sum_{i = {\rm intersection\ of}\ S} sgn(i)\ n_1 n_2
\end{equation}
where $n_1$ and $n_2$ are the $U(1)$ representations carried by the
intersecting pieces of $S$,
and $sgn(i)$ is the sign of the intersection.\footnote{
If coordinates $\sigma_1^a$ and $\sigma_2^a$ are defined on the intersecting
patches $1$ and $2$ of
$S$ which are right handed according to the orientations of these patches then
$sgn(i) = +1$ if
$(\sigma_1^1,\sigma_1^2,\sigma_2^1,\sigma_2^2)$ forms righthanded coordinates
on spacetime, and
$sgn(i) = -1$ if these are left-handed.} See Fig.
Note that $n \rightarrow -n$ if the orientation of $S$ is reversed.

The F~-~F dual theory is a (somewhat trivial) diffeo invariant
theory. Our expectation that
world sheet actions of diffeo invariant theories depend on the diffeo
equivalence class of
the world sheet only is confirmed in the case of this theory. It is not so easy
check, however,
that the sum over worldsheets is diffeo invariant.

The third $U(1)$ theory the author has examined might be called $U(1)$ gravity.
It is diffeo invariant, but it is not a topological theory, having an infinite
dimensional
state space. Its action is
the closest $U(1)$ analogue of the Plebanski action for GR. The Plebanski
action for GR is
\cite{Plebanski77}\cite{CDJ}\cite{Peldan93}:
\begin{equation}\label{plebanski_I}
I[\Sigma,A] = \int \Sigma_a\wedge F^a + \phi^{ab}\Sigma_a\wedge\Sigma_b
\end{equation}
where
$A$ is the left-handed spin connection (which takes its values in complexified
$su(2)$),
$F$ is its curvature, and $\Sigma$ is an $su(2)$ valued 2-form. $\phi^{ab}$ is
a symmetric, trace free,
matrix in the $su(2)$ indices, which acts as a lagrange multiplier.
The $U(1)$ gravity action is obtained from (\ref{plebanski_I}) by dropping the
$su(2)$
indices:\footnote{$U(1)$ gravity is really most closely related to the
Husain-Kuchar theory
\cite{HusainKuchar90} obtained from GR by dropping the requirement in
(\ref{plebanski_I}) that
the lagrange multiplier $\phi$ be traceless.}
\begin{equation}\label{U1gravity}
I[\Sigma,A] = \int \Sigma\wedge F + \phi\Sigma\wedge\Sigma.
\end{equation}

The worldsheet form of this theory is a little different from that of the
preceding theories.
Integrating out the lagrange multiplier produces a delta function in the sum
over surfaces
(\ref{surf_sum}) that restricts this sum to surfaces satifying a condition at
intersections.
Namely, at an intersection of $r$ parts of $S$
\begin{equation}
\sum_{a,b} sgn(i_{a,b}) n_a n_b = 0
\end{equation}
where $a, b\in \{1,...,r\}$ label the intersecting parts of
$S$.\footnote{The simplest allowed intersection is of three surfaces, with $n_1
= n_2 = 2$,
$n_3 = 1$ and appropriately chosen orientations.}
On allowed surfaces the action is identically
zero. Thus the evolution operator is given by a democratic sum over a
restricted class of
allowed surfaces. Note that, as expected, the restriction is diffeo invariant.
Possibly the
evolution operator for GR is given by a similar sum over a restricted class of
surfaces.

Besides $U(1)$ theories the author has also worked out the worldsheet
formulation
of lattice $SU(2)$ BF theory and YM theory.

The connection action for BF theory is
\begin{equation}\label{BFaction}
I_c = \int tr[B\wedge F]
\end{equation}
where $B$ is an $su(2)$ valued
2-form, like $\Sigma$ in the Plebanski action (\ref{plebanski_I}).
(\ref{BFaction}) is just the
Plebanski action without the lagrange multiplier term.

In the lattice, link formulation BF theory has the Feynman weight
\begin{equation}
e^{i\,I} = \sum_{\{j_p\}}\prod_{p\in{\rm plaquettes}} tr[
\prod^{\rightarrow}_{l\in\di p} U^{(j_p)}
(g_l)]   \label{BF_linkweight}
\end{equation}
$\prod^{\rightarrow}_{l\in\di p}$ is the path ordered product around $\di p$,
so
$tr[ \prod^{\rightarrow}_{l\in\di p} U^{(j_p)}(g_l)] = W_{\di p}^{(j_p)}$. The
Wilson loop
$W_{\di p}^{(j_p)}$ can also be written as a trace of $f_p$, so
(\ref{BF_linkweight})
is of the form (\ref{jp_decomposition}).

\begin{figure}
\centerline{\psfig{figure=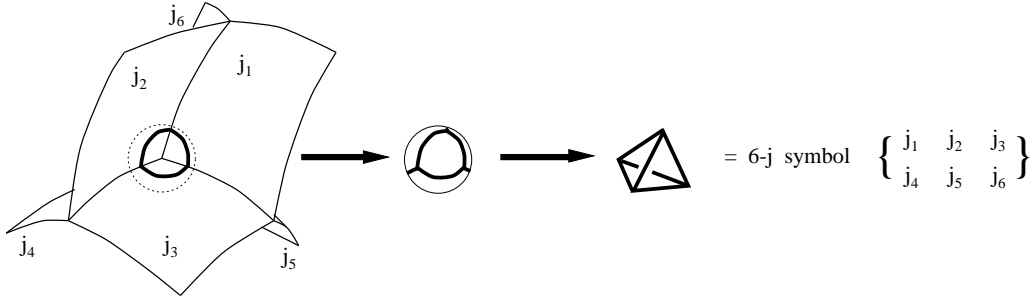,height=4cm}}
\caption{The evaluation of the intersection factor associated with the meeting
of four trivalent
lines of intersection in $2 + 1$ GR (= $2 + 1$ BF theory) is shown. A sphere is
placed around
the intersection point, and the graph on which the intersecting surfaces cut
the sphere is
extracted. This is just a tetrahedron with the spins of the surfaces labelling
the corresponding
edges. This tetrahedron is then evaluated as a 6-j symbol.}
\label{intersection}
\end{figure}

The Feynman weight of a worldsheet, $S$, in the colored surface formulation of
this theory is
obtained by applying the machinery of the last section to
(\ref{BF_linkweight}). It is given by a
factor $(2j + 1)^\chi$ for each simple component of $S$, $\chi$ being the Euler
charactersitic
of the component, and a Racah-Wigner symbol for each point of intersection
(where several lines
of intersection meet).
\begin{equation}\label{BFweight}
e^{i\,I(S)} = \prod_{\sigma\in\mbox{\scriptsize simple components}}(2j_\sigma +
1)^{\chi_\sigma}
				\prod_{\mbox{\scriptsize intersection points}} \mbox{R-W symbol}
\end{equation}
The Racah-Wigner symbol associated with an intersection is computed by placing
a small 3-sphere
around the intersection point, so that each surface meeting at the intersection
point
cuts the sphere along a line (including, of course, the 2-cells of the tree
surface used to expand lines of intersection of valency $r>3$). These lines
form a trivalent graph
labeled by the $j$'s of the surfaces (of the same form as the graphs labeling
the graph basis,
when their vertices are expanded into trivalent trees). The Racah-Wigner symbol
associated with
the intersection is the evaluation, according to the rules of \cite{Yutsis62},
of this graph.
That is, it is the value of the link rep. graph wavefuction evaluated on a flat
connection on the
sphere.
 Fig \ref{intersection}
illustrates this the evaluation of intersection factors.

The resulting theory is easily shown to be equivalent to Ooguri's
\cite{Ooguri92} simplicial formulation of BF
theory. $2 + 1$ dimensional GR leads to exactly the same worldsheet action, and
the worldsheet
formulation is easily seen to be equivalent to the Ponzano-Regge
\cite{Ponzano68} form of GR. In
fact, Turaev and Viro \cite{TV92}, and Kauffman and Lins \cite{Kauffman94} have
given a
formulation of $2+1$ GR virtually identical
to the colored worldsheet form, but with the more general structure $SU(2)_q$
replacing
$SU(2)$.

Note that, in BF theory, surfaces that intersect transversely (i.e. at a point)
do not interact.
Our experience with ``$U(1)$ gravity" suggests that this is due to the absence
of a $B\wedge B$
term in the action, and that in GR such surfaces will interact.

The lack of interaction of surfaces at transverse intersection points in BF
theory
leads the author to suspect that, in the elementary surface formulation of BF
theory, surfaces do not interact at all, the complicated intersection factors
in the colored
surface formulation coming from counting distinct routings of elementary
surfaces. Iwasaki \cite{Iwasaki94} has formulated an elementary surface form of
$2+1$
GR, but his result is too implicit to allow an easy check of this conjecture.

Euclidean $SU(2)$ YM theory can be defined in the continuum connection
formulation by
the Feynman weight
\begin{equation}
e^{i\,I} = \int {\cal D}B\ \  e^{i\int tr[B\wedge F]}\  e^{-g^2 \int
tr[ B_{\alpha\beta} B_{\gamma\delta}]g^{\alpha\gamma} g^{\beta\delta}
\sqrt{g}\epsilon},
\end{equation}
like that of BF theory, except for the second factor in the integrand. $g^2$ is
the YM coupling
strength.

A very nice form of YM theory on a hypercubic lattice is provided by the heat
kernel action
\cite{Menotti81}
\begin{equation}\label{YM_latticeweight}
e^{i\,I} = \sum_{\{j_p\}}\prod_{p\in{\rm plaquettes}} tr[
\prod^{\rightarrow}_{l\in\di p} U^{(j_p)}
(g_l)]\ e^{-g^2\,j_p(j_p+1)},
\end{equation}
again, just like BF theory with an extra factor in the summand.
(\ref{YM_latticeweight}) then leads
to a colored worldsheet Feynman weight like that of BF theory, except that each
simple component
contributes an extra factor $e^{-j(j+1)(\mbox{area})/a^2}$
($a$ = lattice spacing), in close analogy with euclidean compact
EM. This worldsheet formulation of lattice YM theory is, in fact, just
a systematization of the strong coupling expansion of Wilson \cite{Wilson74},
applied to the heat kernel action. See \cite{Itzykson89}.

An elemetary surface formulation of $SU(2)$ YM theory would
constitute a generalization to a 4-dimensional, lattice of
the string representations of 2-d YM theory of Gross and Taylor
\cite{GT93a} \cite{GT93b}.

\section*{Acknowledgments}

The work described here was done while I was at the Department of Physics of
Washington University,
in St. Louis. I thank Clifford Will for his patience.

Discussions with Abhay Ashtekar, Jayashree Balakrishna, Maarten Golterman, Lou
Kauffman,
Jorma Louko, Peter Meisinger, Michael Ogilvie, Lee Smolin, Malcolm Tobias and
Matt Visser have been helpful.

I also thank Xiao Feng Cai.

%\bibliographystyle{alpha}
%\bibliography{loopsbib}

\end{document}